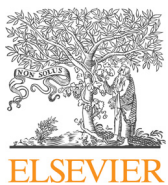
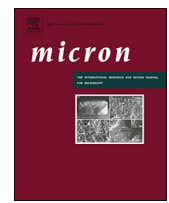

# Automatic microscopic image analysis by moving window local Fourier Transform and Machine Learning

Benedykt R. Jany*, Arkadiusz Janas, Franciszek Krok

*The Marian Smoluchowski Institute of Physics, The Jagiellonian University, Lojasiewicza 11, PL-30348 Krakow, Poland*

ABSTRACT

Analysis of microscope images is a tedious work which requires patience and time, usually done manually by the microscopist after data collection. The results obtained in such a way might be biased by the human who performed the analysis. Here we introduce an approach of automatic image analysis, which is based on locally applied Fourier Transform and Machine Learning methods. In this approach, a whole image is scanned by a local moving window with defined size and the 2D Fourier Transform is calculated for each window. Then, all the Local Fourier Transforms are fed into Machine Learning processing. Firstly, a number of components in the data is estimated from Principal Component Analysis (PCA) Scree Plot performed on the data. Secondly, the data are decomposed blindly by Non-Negative Matrix Factorization (NMF) into interpretable spatial maps (loadings) and corresponding Fourier Transforms (factors). As a result, the microscopic image is analyzed and the features on the image are automatically discovered, based on the local changes in Fourier Transform, without human bias. The user selects only a size and movement of the scanning local window which defines the final analysis resolution. This automatic approach was successfully applied to analysis of various microscopic images with and without local periodicity i.e. atomically resolved High Angle Annular Dark Field (HAADF) Scanning Transmission Electron Microscopy (STEM) image of Au nanoisland of fcc and Au hcp phases, Scanning Tunneling Microscopy (STM) image of Au-induced reconstruction on Ge(001) surface, Scanning Electron Microscopy (SEM) image of metallic nanoclusters grown on GaSb surface, and Fluorescence microscopy image of HeLa cell line of cervical cancer. The proposed approach could be used to automatically analyze the local structure of microscopic images within a time of about a minute for a single image on a modern desktop/notebook computer and it is freely available as a Python analysis notebook and Python program for batch processing.

## 1. Introduction

Nowadays, due to digital data acquisition and storage, a huge amount of data is collected during microscopic imaging in the form of digital images from a single sample. These images are later analyzed by the microscopist to extract the information about a local structure of the measured samples. The analysis requires practice and patience and it is usually tedious work for microscopist which requires a lot of time. The results of the analysis might be biased by the human who performs the analysis by introducing hardly controllable systematic effects in the final results. So far, other approaches were developed to automate the analysis of local structures on microscopic images and usually they are specialized in a particular image type. For identification of local periodic structures in biological samples plug-ins for ImageJ/FIJI (Schindelin et al., 2012) software package like Image-Directionality (Liu, 1991) and OrientationJ (Püspöki et al., 2016) were developed.

They can be used for analysis of preferred orientation and isotropy properties of structures present on an image. The High Resolution Transmission Electron Microscopy (HRTEM) images could be analyzed by Fast Fourier Transform (FFT) to determine a local displacement and strain fields (Hÿtch et al., 1998). This is done by Geometric Phase Analysis (GPA) which uses two non-collinear Fourier phase components of the complex image to derive local displacement and strain fields in two dimensions. For an analysis of local crystallinity from HRTEM fringe pattern approach based on sliding FFT was recently developed (Alxneit, 2018). It uses local FFT which later is fitted by the Gaussian to extract local lattice spacing and direction maps. There are also different software approaches for an analysis of local structure of atomically resolved STEM images (De Backer et al., 2016; Nord et al., 2017). The STEM atomic columns are approximated by two dimensional Gaussian model what requires a lot of computing power especially for large images. The atomic columns intensity and other parameters like ellipticity are extracted. Later, usually some statistical approach is used for intensity quantification. Recently, for the atomically resolved images approaches based on local crystallography and Machine Learning were developed (Belianinov et al., 2015; Vasudevan et al., 2015). They extract local information on material structure based on statistical analysis of atomic neighborhoods based on Fourier Transform followed by clustering and multivariate algorithms like Principal Component Analysis (PCA), Independent Component Analysis (ICA). The Machine Learning approaches could be also successfully used for hyper-spectral EDX imaging in STEM (Rossouw et al., 2015) and in SEM (Jany et al., 2017b). The collected hyperspectral images are in form of data cube i.e. for each spatial pixel a full EDX spectrum is recorded. In following, the data are decomposed by Machine Learning Blind Source Separation (BSS) which allows Energy Dispersive X-ray Spectroscopy (EDX) quantification of nanoheterostructures in STEM or real EDX quantification at nano-scale from simple SEM EDX. The Machine Learning approach is also used for scanning precession electron diffraction (SPED) imaging (Martineau et al., 2019). In SPED a sample is scanned by focused electron beam and for each spatial pixel a full 2D diffraction is recorded, thus, forming 4D data set i.e. two spatial dimensions related to scanning area and two reciprocal related to the diffraction. Later, for this 4D data set the Machine Learning matrix decomposition methods are applied like NMF which can successfully extract local structure features, unmix and reduce dimensionality of complicated data.

* Corresponding author.
 *E-mail address:* benedykt.jany@uj.edu.pl (B.R. Jany).

Here, we present a method based on a local moving window Fourier Transform and Machine Learning blind decomposition of the data via Non-Negative Matrix Factorization (NMF) (Smaragdis et al., 2014), which results in the successful automatic analysis of the local structure of various microscopic images within a minute on a standard notebook or desktop computer, without any external input. It is worth to notice that our approach in contrast to Neutral Network based approaches (e.g. Deep Learning) does not require any training datasets.

## 2. The idea of the method

The idea of the presented method is graphically shown in Fig. 1. It consists of several following steps:

1. First, the microscopic image in form of gray scale tiff file or color png, jpg, or bmp file is read by the program as the Numpy array (Walt et al., 2011). Next, in the image that is a subject of analysis, the square local window with a side size equal to *elemensize* (default value 128 pixels) is defined. This selects the local region of interest (ROI) in the image. The local window is next translated in the image in x and y direction by *xstep* (default value 64 pixels) and *ystep* (default value 64 pixels) appropriately, so the whole image is covered, see Fig. 1 (left). The choice of *xstep* and *ystep* defines the resolution of the performed image analysis and also it influences the computational time since the number of generated ROIs is inversely proportional to the product of *xstep* and *ystep*. The finer the resolution is, the more ROIs are generated for the analysis. While the *choice of elemensize* (Window Size) defines not only the resolution but also the sensitivity scale of the analysis to local structures variation inside the window. The selected default values are an optimal matter of choice for microscopic images of a size of $\sim 2000 \times 2000$ pixels, which are typically nowadays collected. The user is advised to use it as a starting point. The program has the option to scale the image width to 2048 pixels. This is very useful for applying program default parameters to an image of arbitrary size.

2. For each local window (ROI) a 2D Power Spectrum Fourier Transform is calculated. To prevent high-frequency noise Hanning window is used in the Fourier Transform calculation. The Hanning windows is commonly used in FFT calculations for high-frequency noise suppression since it has a small impact on frequency resolution and amplitude accuracy comparing to other windowing functions. The computed ROIs are multiplied by the Hanning windows function. Thus, after this step, the data consist of two spatial dimensions X, Y in real space, as defined by the image, and corresponding two dimensions in reciprocal space $K_X$, $K_Y$, as defined by 2D Power Spectrum for each local window. The output of this step is presented in Fig. 1 (center). It is seen that the resolution of the image is decreased, since now one pixel corresponds to the size of [*xstep*, *ystep*] which is [64 pixels, 64 pixels], as for default parameters. This together forms 4D data set very similarly to scanning precession electron diffraction (SPED) 4D data (Martineau et al., 2019).

3. Next, the formed 4D data set is processed by Machine Learning methods. To determine the number of components/fractions present in the data, the Principal Component Analysis (PCA) is performed. Here we used the implementation from Scikit-Learn library (Pedregosa et al., 2011) as in Hyperspy toolbox (Peña et al., 2017). From the PCA Scree Plot (portion of variance vs. component number) by identifying the inflection point (a point at which a change in curvature occurs) the number of components in the data is found, i.e., the components which exhibit significantly higher variance, see Fig. 1 (right). Our program has the possibility to automatically find candidates for the inflection points from the Scree Plot. This is done by numerically finding local maxima of the gradient of the Scree Plot. The user can inspect and use the automatically found inflection point value as first guess of the number of components. Next the data are decomposed blindly, without any external inputs concerning the image, using only number of components as found by PCA Scree Plot analysis by Non-Negative Matrix Factorization (NMF) (Smaragdis et al., 2014) which can successfully realize dimensionality reduction and unmixing of the 4D data (Martineau et al., 2019). The NMF decomposition requires only a non negativity of the data and it is, nowadays, widely used in different fields of science as a good and reliable data decomposition method (Berné et al., 2008; Eggeman et al., 2015; Virtanen, 2007). The non-negativity constraint of NMF results in parts based representations of the objects which is different from other decomposition methods like PCA, which provide holistic representations (Lee et al., 1999). The NMF decomposes blindly the data matrix into two non-negative matrices by minimizing the distance function which is an extension of the Euclidean norm to matrices i.e. the Frobenius norm. The matrices are initialized by two non-negative singular value decomposition processes. Here we used the implementation from Scikit-Learn library (Pedregosa et al., 2011) as in Hyperspy toolbox (Peña et al., 2017). As an output of NMF decomposition the interpretable spatial maps (loadings) and corresponding 2D Power Spectra (factors) are generated, see Fig. 1 (right panel). The image is automatically analyzed i.e. the local features are automatically discovered based on FFT which is sensitive to the local orientation and spacing changes. The program outputs a PDF file report with images and two tiff files which store original loadings and factors data. The files can be read and further analyze by free software ImageJ/FIJI (via import Bio-Formats) (Schindelin et al., 2012). The user can now inspect resulted decomposition and decide whenever the higher resolution is desired, this could be accomplished by changing *xstep* and *ystep* to lower values (i.e. 32 or 16 pixels) from default 64 pixels. The user has to keep in mind that by increasing the resolution the computational time will increase since number of generated ROIs is inversely proportional to the product of *xstep* and *ystep*. Changing *xstep* and *ystep* from 64 pixels to 16 pixels induce increase in computing time from about 1 min to about 10 min for a single image analysis.

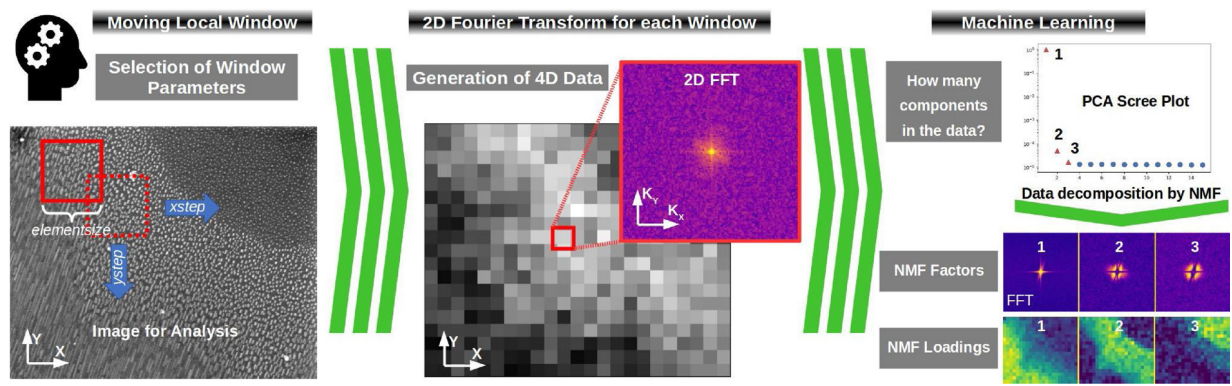

**Fig. 1.** Graphical presentation of the idea of the method. Moving local window is applied to the microscopic image and for each local window a 2D Power Spectrum Fourier Transform is computed resulting in formation of 4D data set. Later the data are fed into Machine Learning processing: firstly, to determine number of components/fractions in the data Principal Component Analysis (PCA) Scree Plot is computed; secondly, the data are decomposed blindly into different interpretable components by Non-Negative Matrix Factorization. The local features in the image are automatically discovered.

The automatic analysis of the microscopic images based on the presented method with a default parameters for the image of $\sim 2000 \times 2000$ pixels takes about a minute on a standard desktop or notebook computer (for the calculations presented in this paper we used Intel i5 CPU @1.6 GHz with four cores and 8GB of RAM). The method is implemented in Python using freely available software library tools as building blocks like Hyperspy (Peña et al., 2017), Scikit-Learn (Pedregosa et al., 2011), Numpy (Walt et al., 2011) and Scipy (Virtanen et al., 2019). The presented method is freely available as Python analysis notebook and python program for batch processing. It could be downloaded from Mendeley Data (Jany, Benedykt R., 2019)

## 3. Applications to various microscopic images and discussion

The proposed approach was successfully applied to various types of microscopic images with well known and defined local features. It was tested on images with a local periodicity present, like atomic columns in HAADF STEM, or reconstruction domains in STM. This approach is also relevant to analyzes of images without a local periodicity present, like surface features in SEM or fibers like structure in Fluorescence microscopy.

### 3.1. Atomically resolved HAADF STEM

As the first example of application of the presented method, an analyses of atomically resolved HAADF STEM image of Au nanoisland grown on reconstructed Ge(001) surface will be shown (see Fig. 2). The island is a result of the thermally induced self-assembly process of deposited thin Au layer on Ge substrate. For this system, when annealed to temperatures above the Au/Ge eutectic temperature, i.e. 634 K, Au/Ge liquid droplet is formed. When the sample is cooled to room temperature, the eutectic liquid recrystallizes, leading to a crystalline Au nanoisland. However, depending on the rate of the cooling process, Au can adopt typical crystalline fcc structure or unusual for it a hcp structure. Thus, the formed final nanoislands are made of two phases of Au, i.e., cubic fcc phase and rare and unusual for Au hexagonal hcp phase separated from each other by defected inter-growth domain along (111) crystallographic planes (Jany et al., 2017a). The image shown in Fig.2 was obtained by Probe Corrected Scanning Transmission Electron Microscope (STEM) FEI Titan at the University of Antwerp, Belgium, operated at 300 kV (Guzzinati et al., 2018). The atomically resolved image in Fig. 2a shows germanium bulk at the bottom, nanoisland in the middle which consists of this two Au phases and platinum capping layer on the top. With the naked eye it is not trivial to distinguish in the island the regions of these two phases of Au. We applied our method with a default parameters, i.e., *elemensize* = 128 pixels and *xstep and ystep* equal to 64 pixels. Then, for the 4D data set generated, first, the PCA analyses was performed and the resulting Scree Plot is shown in Fig. 2b. It is clear that the first seven components exhibit significantly higher variance than the remaining ones. In following, we used NMF to decompose the data assuming only the seven components, as derived from PCA. The NMF analyses results in seven decomposition factors as shown in Fig. 2c (2D FFT Power Spectra) and corresponding decomposition loadings, in Fig. 2d, presenting spatial distribution maps. Examining the 2D FFT Power Spectra factors (Fig. 2c) we can clearly identify the phases/components present in the image. Thus, the component 1 corresponds to the background and it is related to the local intensity variation in the image, which is reflected in the changes of intensity in the low frequency peak at the center in FFT. This is very similar to the TEM Bright Field imaging, where the imaging is performed using only using center spot in the diffraction. This raises in the appearance of the crystallographic-like contrast. In our case in the component 1 also different crystallographic regions are nicely visible. The component 2 shows defined maxima in FFT which correspond to the Germanium. The component 3 FFT shows a structure without well defined maxima and this could be attributed to the structure lacking of well-developed long range order like platinum deposition capping layer or diffused interfaces. The components 4 and 5 FFT of hexagonal-symmetry patterns show well defined maxima which correspond to the two orientations of Au hcp phase. The component 6 FFT of square-symmetry diffraction pattern corresponding to the Au fcc phase. The component 7 FFT corresponds to the defected region of inter-growth domain between Au hcp and Au fcc phases. All phases/components present in the image are correctly automatically discovered by the method without any external input from the user with a time of about a minute. This easily allows for extraction of Au hcp relative phase content and saves hours of manual data analysis time spend by the microscopist. A higher resolution result (xstep = 16 pixels, ystep = 16 pixels) is available in Appendix A Supporting Information page 2.

### 3.2. Scanning tunneling microscopy (STM)

The next studied example is a Scanning Tunneling Microscopy image of the Au-induced c(8 × 2) reconstructed Ge(001) surface. A cleaning procedure for the Ge(001) surface, which proceeds in UHV conditions by ion-beam sputtering and annealing, results in atomically flat terraces of isotropic c(4 × 2) reconstruction. On this substrate surface, after annealing of deposited thin Au layer, domain-like c(8 × 2) reconstruction appears in a form of atomic Au-induced chains. These 1D chains assembled on adjacent terraces are rotated by 90 deg. with respect to each other due to the reconstruction of the initial substrate surface (Krok et al., 2014). The 1D atomic chains exhibit 1D conductivity i.e. Tomonaga-Luttinger liquid-like conductivity (Blumenstein et al., 2011). In Fig.3 STM image of Au-induced nanowires on Ge(001) surface after annealing at 720 K for 15 min is shown. The surface was imaged in situ in UHV conditions (pressure $\sim 10^{-10}$ Pa) by Scanning Tunneling Microscope Omicron RT STM/AFM system. The obtained raw STM images were corrected to piezo scanner distortions by standard procedures using free software Gwyddion (Nečas and Klapetek, 2011) i.e. leveling by mean plane subtraction followed by scanning rows aligning. The image was scaled four times and saved as 16-bit tiff file. The image scaling was performed to obtain the image of ∼ 2000 × 2000 pixels. In following, the presented method, with a default parameters, was applied to analyze the STM image shown in Fig.3a. The 4D data set was generated and with the PCA analyses performed, the Scree Plot was generated as shown in Fig. 3b. From the Scree Plot it is clear that the first 3 components exhibit significantly higher variance as compared to the remaining components. Next, we used NMF to decompose the data assuming only these three components present. The NMF analyses result in the three decomposition factors, shown in Fig. 3c (2D FFT), and corresponding decomposition loadings exhibiting spatial distribution maps as shown in Fig. 3d. Examining the resulted Fourier Transforms we can easily identify the formed structures. The component 1 corresponds to the first region of Au-induced chains (domain 1). The component 2 is identified as terrace edges. The component 3 corresponds to the domain 2 composed of Au-induced chains rotated by 90 deg. with respect to those of domain 1. Thus, these two domains are correctly recognized in the STM image. A higher resolution result (xstep = 16 pixels, ystep = 16 pixels) is available in Appendix A Supporting Information page 2.

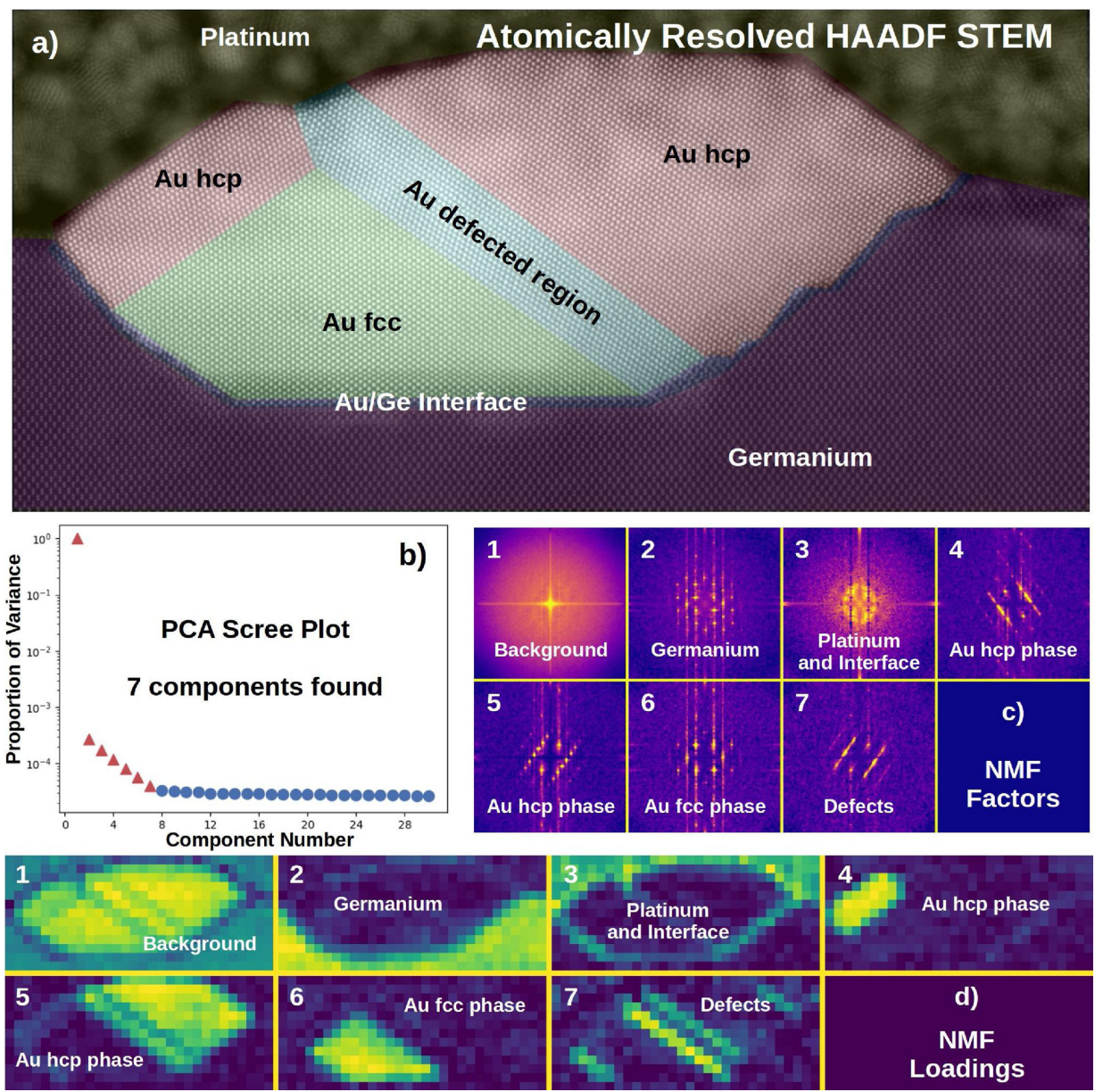

**Fig. 2.** Results of the presented approach applied to atomically resolved HAADF STEM Image (2048 × 960 pixels) of nanoisland of Au fcc/hcp phase grown on Ge (001) with a color overlay of phases a) (Jany et al., 2017a) (CC by 4.0). In b) PCA Scree Plot of first 30 principal components is shown. The first 7 components exhibit significantly higher variance as compared to the others. In c) Non-Negative Matrix Factorization (NMF) decomposition results of moving local window Fourier transform data in form of decomposition factors (here Fourier transforms) are shown and corresponding decomposition loadings (spatial maps) are in d). The image width is equal to 49.7 nm.

### 3.3. Scanning Electron microscopy

Another example is Scanning Electron Microscopy image of Au-rich morphology resulted from the temperature induced self-assembly process of 2 ML of Au deposited on clean GaSb(001) surface. The cleaning procedure of GaSb(001) surface consist of cycles of ion-sputtering and annealing performed in the UHV conditions. In the case of compound samples, here GaSb, this cleaning procedure influences the surface stoichiometry due to preferential sputtering of Sb and leads to enrichment of the clean surface in gallium (Jany et al., 2018). If the enrichment in gallium is not uniform across the substrate surface, in the process of the thermally-induced self-assembly of deposited thin Au layer, one can obtained different nanostructures morphologies, ranging from cluster to elongated nanowires like the ones shown in Fig. 4a. We analyzed this SEM image with default parameters. The PCA Scree Plot is presented in Fig. 4b. For this case, 3 components exhibit significantly higher variance than the remaining components. In following, the NMF analyses was used to decompose the data assuming only the first three components. The NMF in this case results in three decomposition factors, seen in Fig. 4c (2D FFT), and corresponding decomposition loadings shown in Fig. 4d. Examining the resulted Fourier Transforms we can identify and characterize the three areas in the image based on their orientations and characteristic spacing. The component 1 FFT shows highly elongated in one direction structures (19 deg direction with respect to the horizontal direction clockwise) which goes to higher frequencies 18 pixels in FFT corresponding to spacing of 13 nm. This is attributed to the nanowires region Area 1. The component 2 FFT shows a ring patter with a size of 8 pixels in FFT corresponding to spacing of 29 nm elongated in one direction (19 deg direction). It is identified as elongated islands of region Area 2. The component 3 FFT shows a ring patter with a size of 11 pixels in FFT corresponding to spacing of 21 nm, only slightly elongated (19 deg direction), indicating preferential direction in this region which is not directly easy visible from the image. This is attributed to the symmetric clusters region Area 3. The spatial maps Fig. 4d show how these structures are distributed in the image, from it one can exactly see what are the boundaries regions between them and how one region evolves into another. Here, also all the features of the SEM image were correctly fully automatically identify based on nanostructures characteristic spacing and their orientation and could be later characterize by these parameters, although the image local features did not exhibit periodicity like for previous two examples. A higher resolution result (xstep = 16 pixels, ystep = 16 pixels) is available in Appendix A Supporting Information page 3.

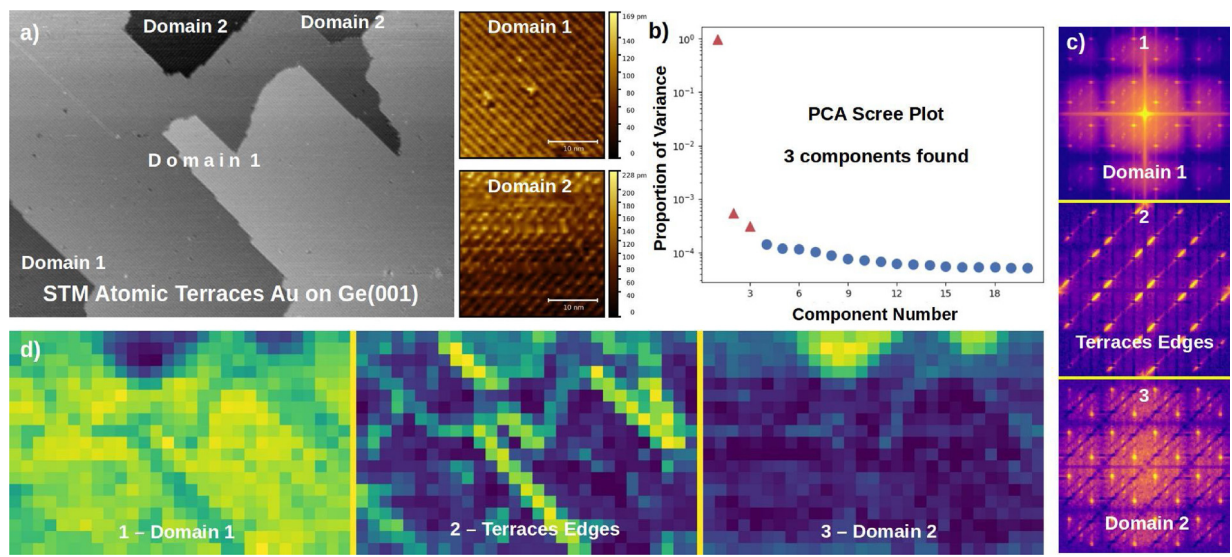

**Fig. 3.** Results of presented approach applied to STM Image (2021 × 1387 pixels) of Au induced reconstruction on Ge(001) which shows atomic terraces with two Ge (001)-c(8 × 2)-Au reconstruction domains a). PCA Scree Plot of first 20 principal components is shown in b); 3 components found which exhibit significantly higher variance. Non-Negative Matrix Factorization (NMF) decomposition results of moving local window Fourier transform data in form of decomposition factors (here Fourier transforms) c) and corresponding decomposition loadings (spatial maps) are in d). The image width is equal to 296 nm.

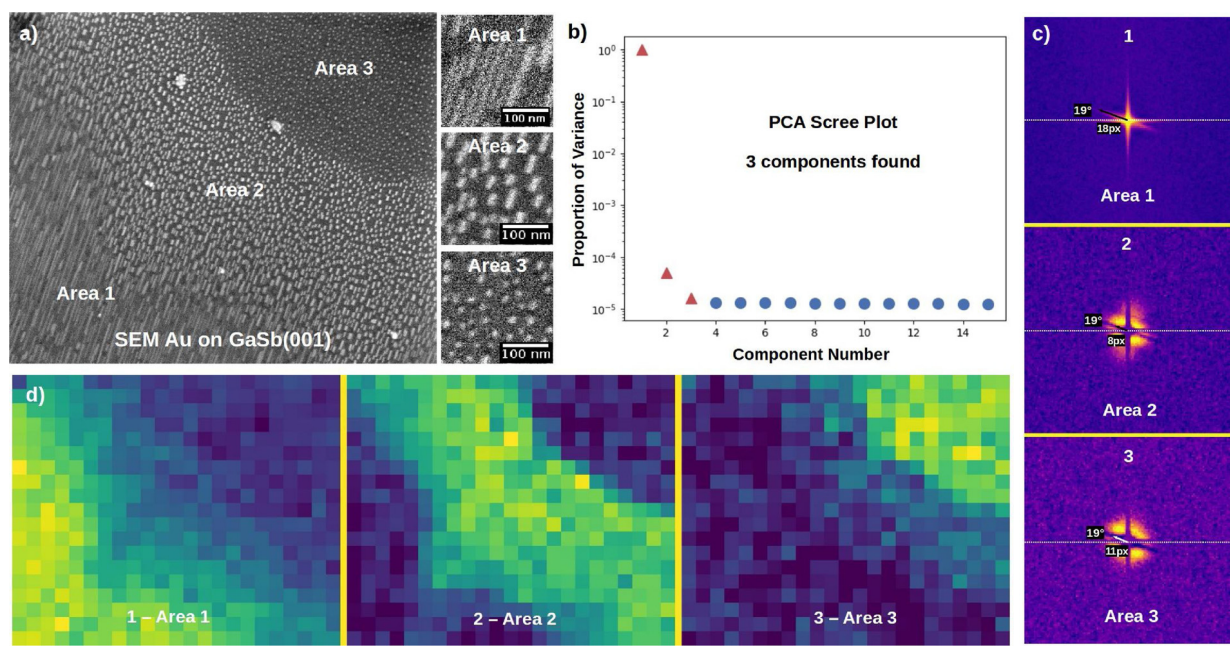

**Fig. 4.** Results of presented approach applied to SEM Image (1662 × 1370 pixels) of Au-rich nanostructures on GaSb(001) surface. The image a) shows areas of three different morphologies: nanowires, elongated islands or symmetric clusters. PCA Scree Plot of first 15 principal components is shown in b). The first 3 components which exhibit significantly higher variance are marked with red. Non-Negative Matrix Factorization (NMF) decomposition results of moving local window Fourier transform data in form of decomposition factors (here Fourier transforms) are in c) and corresponding decomposition loadings (spatial maps) d). The image width is equal to 3 microns. (For interpretation of the references to colour in this figure legend, the reader is referred to the web version of this article.)

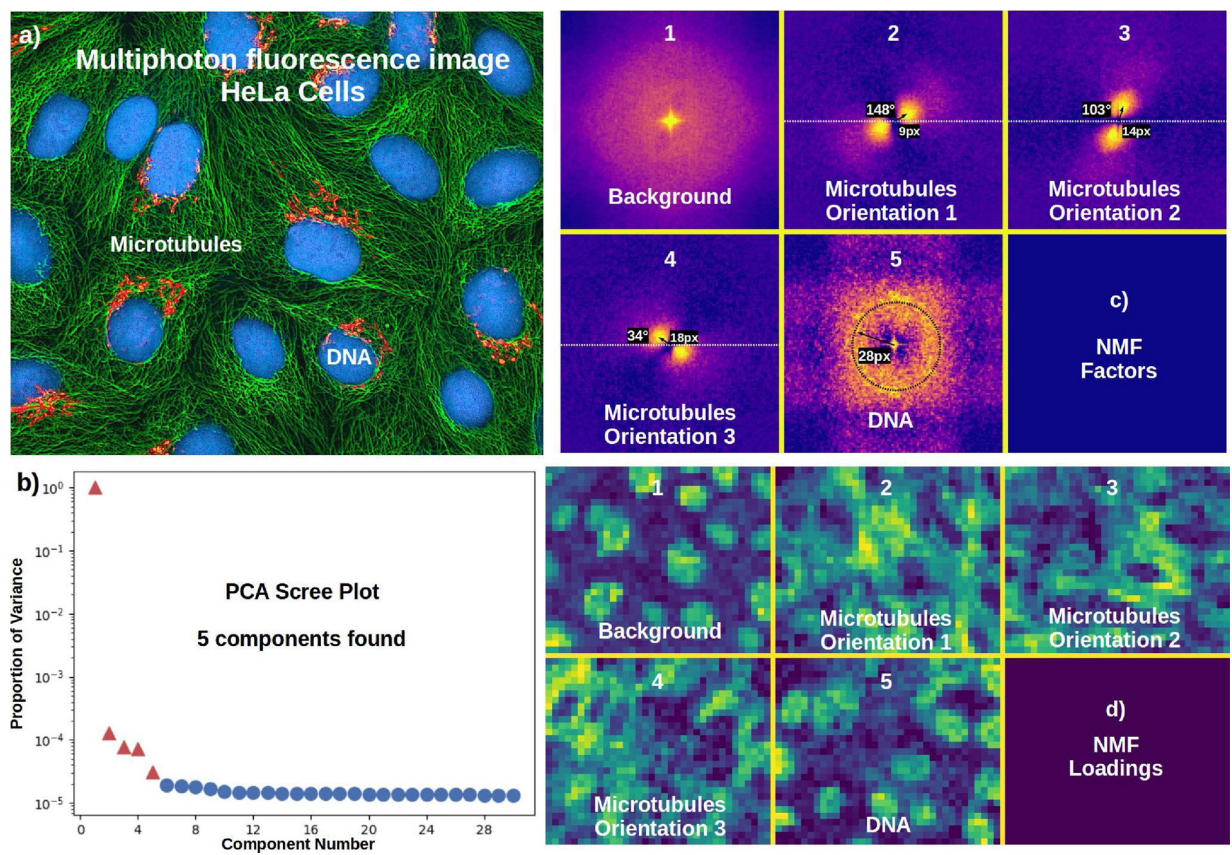

**Fig. 5.** Results of presented approach applied to Multi-photon fluorescence image (2400 × 1999 pixels) of cultured HeLa cell line of cervical cancer with a fluorescent protein targeted to the Golgi apparatus (orange), microtubules (green) and counterstained for DNA (cyan) a). The image is public domain and is from National Institute of Health. One can notice that microtubules fibers like structure exhibits different spatial orientations. PCA Scree Plot of first 30 principal components b); 5 components found which exhibit significantly higher variance. Non-Negative Matrix Factorization (NMF) decomposition results of moving local window Fourier transform data in form of decomposition factors (here Fourier transforms) c) and corresponding decomposition loadings (spatial maps) d). (For interpretation of the references to colour in this figure legend, the reader is referred to the web version of this article.)

### 3.4. Fluorescence microscopy

The last example is a Multi-photon fluorescence microscopy image of cultured HeLa cell line (Lucey et al., 2009) of cervical cancer with a fluorescent protein targeted to the Golgi apparatus (orange), microtubules (green) and counter-stained for DNA (cyan) as shown in Fig. 5a. The Nikon RTS2000MP custom laser scanning microscope was used. One can notice that microtubules fibers-like structures exhibit different spatial orientations. Our approach was applied to this image with default parameters after conversion the image to 32-bit greyscale. First the PCA was performed on 4D data set and the resulted Scree Plot (Fig. 5b) exhibits five components with significantly higher variance. Next, the NMF decomposition was performed assuming this five components. The resulting factors (see in Fig. 5c) and loadings (shown in Fig. 5d) were obtained. By examining the FFT factors we can clearly identify the phases/components present in the image and their properties based on the FFT pattern. The component 1 corresponds to the background and it is related to the intensity variation in the image. The component 2 FFT shows structure with defined preferential orientation in the 148 deg direction (with respect to the horizontal direction clockwise), the structure maximum to center distance in FFT is 9 pixels which corresponds to the characteristic spacing of 14 spatial pixels. This component corresponds to the microtubules fiber-like structure with defined orientation. In the same way, the component 3 corresponds to the microtubules region with preferential orientation in the 103 deg direction and characteristic spacing of 14 spatial pixels. Whereas the component 4 corresponds to the microtubules region with preferential orientation in the 34 deg direction and characteristic spacing of 18 spatial pixels. The component 5 FFT exhibits symmetric ring-like structure, which indicates that no particular direction is preferred, with a radius of 28 pixels in FFT which correspond to the characteristic spacing between the structures of 5 spatial pixels. This components corresponds to the DNA which is distributed within the nucleus of the cell. A higher resolution result (xstep = 16 pixels, ystep = 16 pixels) is available in Appendix A Supporting Information page 3.

Here, once more, the features on the image and their properties as derived from FFT NMF factors (orientation, characteristic spacing) are automatically discovered without any external input within a time of about a minute.

The presented analysis approach could be also successfully applied to other than microscopic images like photography or a painting, to analyze local features on them. For examples see Appendix A Supporting Information page 4–9.

For all above analysis we used default program parameters with *elemensize = 128 pixels* (moving window size), the choice *of elemensize* defines not only the resolution but also the sensitivity scale of the analysis to local structures variation inside the window. One may think about it as a local eye, which only sees the features inside the window. To understand this lets consider example of the atomically resolved image, which consists of atomic structure of some crystalline material. The atomic structure consists of building blocks i.e. unit cells, which by translation form crystalline material. If one selects here the *elementsize* in the order of the unit cell size, then our local eye will see the crystalline structure when moved over the image. But when we select the *elementsize* smaller than the unit cell size, then our local eye which moves over the image will never see the crystalline structure. The same happens when *elementsize* is very big, then our local eye will no longer see the local unit cell variations from one place to another only a global view. The selection of *elementsize* defines the local sensitivity scale of the performed analysis and is the major source of the analysis error, since wrong selection of it will result in the analysis which will lack some image local feature detection. To address this problem the user is advised to check how the size of *elementsize* influences the number of PCA components. Example of such check is presented in Fig. 6, which shows estimated number of components from PCA Scree Plot analysis as a function of *elementsize* (moving window size) in pixels for analyzed images in this paper. It is seen that for very small (16 pixels) and very big (512 pixels) *elementsize* number of PCA components is either 1 or 2, only limited number of local features is detected. It is seen also that, the presented dependences exhibit a maxima which corresponds to the optimal value of *elementsize* where all local features are detected, this particularly can be for several values of *elementsize*. It is seen that for all analyzed images in this paper *elementsize = 128 pixels* (default value) seems to be optimal i.e. provides the highest number of components for analyzed images in this paper.

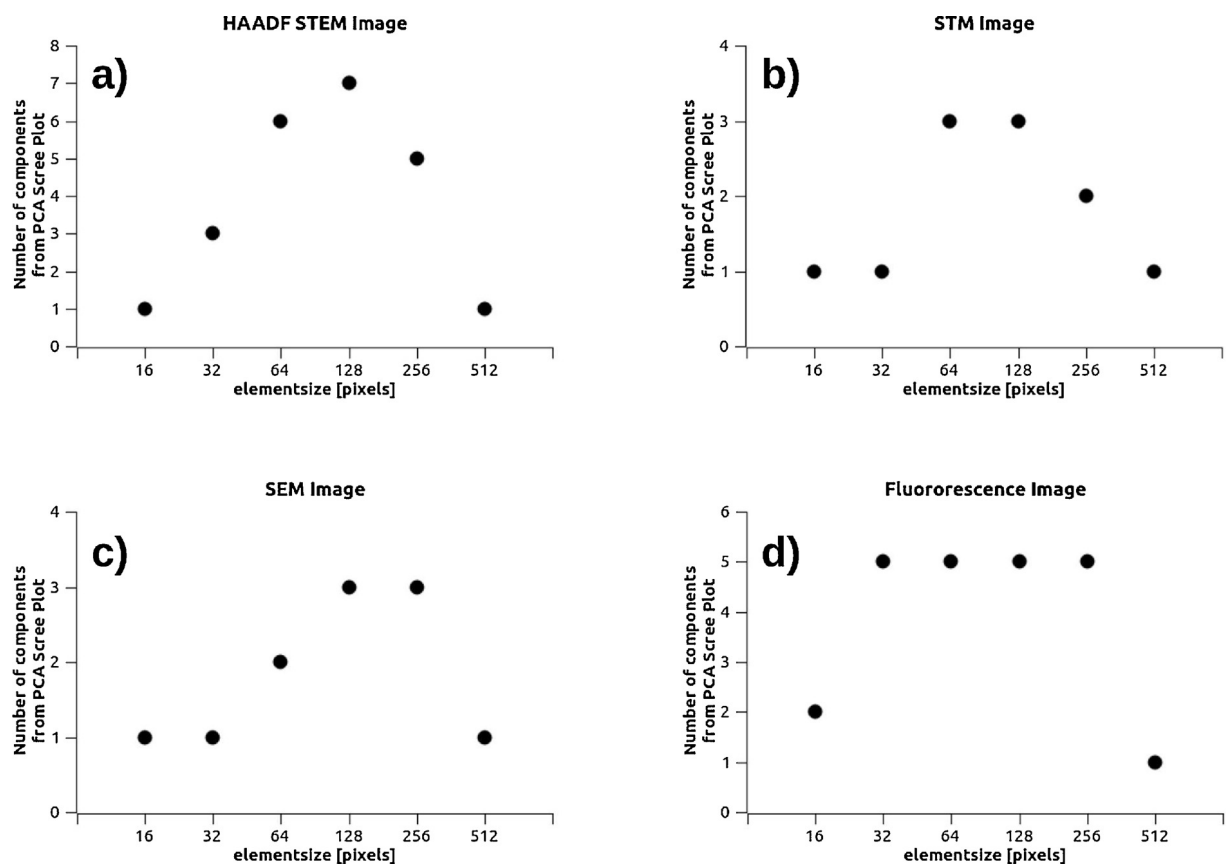

**Fig. 6.** Optimizing local feature detection error. Estimated number of components from PCA Scree Plot analysis as a function of elementsize (moving window size) in pixels for analyzed images. Too small or too big selection of elementsize results in small number of components, not all local features are detected. It is seen that for all analyzed images elementsize = 128 pixels seems to be optimal i.e. provides the highest number of components.

## 4. Conclusions

We have presented approach of an automatic analysis of microscopic images based on moving local window FFT. This approach is sensitive to the local features of the image like symmetry, orientation, characteristic spacing. The generated 4D data set of local FFT's is later decomposed blindly using Machine Learning approach Non-Negative Matrix Factorization (NMF), where number of decomposition components is obtained by the user from previously applied PCA Scree Plot analysis. Our method successfully analyzed different microscopic images with and without periodicity present. As it was shown it was successfully applied to the atomically resolved HAADF STEM image, STM image, SEM image and Fluorescence image. All the features on the analyzed images were automatically discovered and later characterized by their Fourier Transform. Additionally, the automatic features discovery by our method has an advantage that the obtained results are not biased by the microscopist which performs the analysis. The program treats each image in the same way i.e. same bias. We showed that the approach could be also applied to other than microscopic images like photography or a painting. It is also worth to notice, that the analysis of each image with a typical size of $\sim 2000 \times 2000$ pixels with default parameters takes about a minute on a standard desktop or notebook computer. This gives a possibility to analyze big amount of image data, obtained during measurement sessions on the microscopes, fully automatically within a reasonable time and computing resources, saving precious microscopist time. Our approach is freely available as a Python analysis notebook and Python program for batch processing.

**Declaration of Competing Interest**

The authors declare no competing financial interests.

**Acknowledgments**

The authors gratefully acknowledges Prof. Johan Verbeeck and Dr. Nicolas Gauquelin from EMAT University of Antwert, Belgium for atomically resolved HAADF STEM imaging of nanoislands made of Au fcc and Au hcp phases. Support by the Polish National Science Center (UMO- 2018/29/B/ST5/01406) is also acknowledged.

**Appendix A. Supporting Information**

Supporting Information related to this article can be found, in the online version, at doi:https://doi.org/10.1016/j.micron.2019.102800.

**References**

Alxneit, I., 2018. HRTEMFringeAnalyzer a free python module for an automated analysis of fringe pattern in transmission electron micrographs. J. Microsc. 271, 62–68. https://doi.org/10.1111/jmi.12695.

Belianinov, A., He, Q., Kravchenko, M., Jesse, S., Borisevich, A., Kalinin, S.V., 2015. Identification of phases, symmetries and defects through local crystallography. Nat. Commun. 6, 7801. https://doi.org/10.1038/ncomms8801.

Berné, O., Joblin, C., Rapacioli, M., Thomas, J., Cuillandre, J.-C., Deville, Y., 2008. Extended Red Emission and the evolution of carbonaceous nanograins in NGC 7023. AA 479, L41–L44. https://doi.org/10.1051/0004-6361:20079158.

Blumenstein, C., Schafer, J., Mietke, S., Meyer, S., Dollinger, A., Lochner, M., Cui, X.Y., Patthey, L., Matzdorf, R., Claessen, R., 2011. Atomically controlled quantum chains hosting a Tomonaga-Luttinger liquid. Nat. Phys. 7, 776–780.

De Backer, A., van den Bos, K.H.W., Van den Broek, W., Sijbers, J., Van Aert, S., 2016. StatSTEM: an efficient approach for accurate and precise model-based quantification of atomic resolution electron microscopy images. Ultramicroscopy 171, 104–116. https://doi.org/10.1016/j.ultramic.2016.08.018.

Eggeman, A.S., Krakow, R., Midgley, P.A., 2015. Scanning precession electron tomography for three-dimensional nanoscale orientation imaging and crystallographic analysis. Nat. Commun. 6, 7267.

Guzzinati, G., Altantzis, T., Batuk, M., De Backer, A., Lumbeeck, G., Samaee, V., Batuk, D., Idrissi, H., Hadermann, J., Van Aert, S., Schryvers, D., Verbeeck, J., Bals, S., 2018. Recent Advances in Transmission Electron Microscopy for Materials Science at the EMAT Lab of the University of Antwerp. Materials 11, 1304. https://doi.org/10.3390/ma11081304.

Hÿtch, M.J., Snoeck, E., Kilaas, R., 1998. Quantitative measurement of displacement and strain fields from HREM micrographs. Ultramicroscopy 74, 131–146. https://doi.org/10.1016/S0304-3991(98)00035-7.

Jany, B.R., 2019. Python jupyter notebook to perform automatic microscopic image analysis by moving window local fourier transform and machine learning. Mendeley Data v2. https://doi.org/10.17632/25x46xjyr5.2.

Jany, B.R., Gauquelin, N., Willhammar, T., Nikiel, M., van den Bos, K.H.W., Janas, A., Szajna, K., Verbeeck, J., Van Aert, S., Van Tendeloo, G., Krok, F., 2017a. Controlled growth of hexagonal gold nanostructures during thermally induced self-assembling on Ge(001) surface. Sci. Rep. 7, 42420.

Jany, B.R., Janas, A., Krok, F., 2017b. Retrieving the quantitative chemical information at nanoscale from scanning Electron microscope energy dispersive X-ray measurements by machine learning. Nano Lett. https://doi.org/10.1021/acs.nanolett.7b01789.

Jany, B.R., Janas, A., Piskorz, W., Szajna, K., Kryshtal, A., Cempura, G., Indyka, P., Kruk, A., Czyrska-Filemonowicz, A., Krok, F., 2018. Chemically driven growth of Au-rich nanostructures on AIII-BV semiconductor surfaces. arXiv:1811.02488 [cond-mat].

Krok, F., Kaspers, M.R., Bernhart, A.M., Nikiel, M., Jany, B.R., Indyka, P., Wojtaszek, M., Möller, R., Bobisch, C.A., 2014. Probing the electronic transport on the reconstructed Au/Ge(001) surface. Beilstein J. Nanotechnol. 5, 1463–1471. https://doi.org/10.3762/bjnano.5.159.

Lee, Daniel, D., Seung, H., Sebastian, 1999. Learning the parts of objects by non-negative matrix factorization. Nature 401, 788–791.

Liu, Z.-Q., 1991. Scale space approach to directional analysis of images. Appl. Opt., AO 30, 1369–1373. https://doi.org/10.1364/AO.30.001369.

Lucey, B.P., Nelson-Rees, W.A., Hutchins, G.M., 2009. Henrietta Lacks, HeLa cells, and cell culture contamination. Arch. Pathol. Lab. Med. 133, 1463–1467. https://doi.org/10.1043/1543-2165-133.9.1463.

Martineau, B.H., Johnstone, D.N., van Helvoort, A.T.J., Midgley, P.A., Eggeman, A.S., 2019. Unsupervised machine learning applied to scanning precession electron diffraction data. Adv. Struct. Chem. Imaging 5, 3. https://doi.org/10.1186/s40679-019-0063-3.

Nečas, D., Klapetek, P., 2011. Gwyddion: an open-source software for SPM data analysis. Open Phys. 10, 181–188. https://doi.org/10.2478/s11534-011-0096-2.

Nord, M., Vullum, P.E., MacLaren, I., Tybell, T., Holmestad, R., 2017. Atomap: a new software tool for the automated analysis of atomic resolution images using two-dimensional Gaussian fitting. Adv. Struct. Chem. Imaging 3, 9.

Pedregosa, F., Varoquaux, G., Gramfort, A., Michel, V., Thirion, B., Grisel, O., Blondel, M., Prettenhofer, P., Weiss, R., Dubourg, V., Vanderplas, J., Passos, A., Cournapeau, D., Brucher, M., Perrot, M., Duchesnay, E., 2011. Scikit-learn: machine learning in Python. J. Mach. Learn. Res. 12, 2825–2830.

Peña, F., Ostasevicius, T., Fauske, V.T., Burdet, P., Jokubauskas, P., Nord, M., Sarahan, M., Johnstone, D.N., Prestat, E., Taillon, J., Caron, J., Furnival, T., MacArthur, K.E., Eljarrat, A., Mazzucco, S., Migunov, V., Aarholt, T., Walls, M., Winkler, F., Martineau, B., Donval, G., Hoglund, E.R., Zagonel, L.F., Garmannslund, A., Gohlke, C., iygr, Chang, H.-W., 2017. hyperspy/hyperspy v1.2. https://doi.org/10.5281/zenodo.345099.

Püspöki, Z., Storath, M., Sage, D., Unser, M., 2016. Transforms and operators for directional bioimage analysis: a survey. In: De Vos, W.H., Munck, S., Timmermans, J.-P. (Eds.), Focus on Bio-Image Informatics, Advances in Anatomy, Embryology and Cell Biology. Springer International Publishing, Cham, pp. 69–93. https://doi.org/10.1007/978-3-319-28549-8_3.

Rossouw, D., Burdet, P., de la Peña, F., Ducati, C., Knappett, B.R., Wheatley, A.E.H., Midgley, P.A., 2015. Multicomponent signal unmixing from Nanoheterostructures: overcoming the traditional challenges of nanoscale X-ray analysis via machine learning. Nano Lett. 15, 2716–2720. https://doi.org/10.1021/acs.nanolett.5b00449.

Schindelin, J., Arganda-Carreras, I., Frise, E., Kaynig, V., Longair, M., Pietzsch, T., Preibisch, S., Rueden, C., Saalfeld, S., Schmid, B., Tinevez, J.-Y., White, D.J., Hartenstein, V., Eliceiri, K., Tomancak, P., Cardona, A., 2012. Fiji: an open-source platform for biological-image analysis. Nat. Methods 9, 676–682. https://doi.org/10.1038/nmeth.2019.

Smaragdis, P., Fevotte, C., Mysore, G.J., Mohammadiha, N., Hoffman, M., 2014. Static and dynamic source separation using nonnegative factorizations: a unified view. IEEE Signal Process. Mag. 31, 66–75. https://doi.org/10.1109/MSP.2013.2297715.

Vasudevan, R.K., Belianinov, A., Gianfrancesco, A.G., Baddorf, A.P., Tselev, A., Kalinin, S.V., Jesse, S., 2015. Big data in reciprocal space: sliding fast Fourier transforms for determining periodicity. Appl. Phys. Lett. 106, 091601. https://doi.org/10.1063/1.4914016.

Virtanen, P., Gommers, R., Oliphant, T.E., Haberland, M., Reddy, T., Cournapeau, D., Burovski, E., Peterson, P., Weckesser, W., Bright, J., van der Walt, S.J., Brett, M., Wilson, J., Jarrod Millman, K., Mayorov, N., Nelson, A.R.J., Jones, E., Kern, R., Larson, E., Carey, C., Polat, İ., Feng, Y., Moore, E.W., Vand erPlas, J., Laxalde, D., Perktold, J., Cimrman, R., Henriksen, I., Quintero, E.A., Harris, C.R., Archibald, A.M., Ribeiro, A.H., Pedregosa, F., van Mulbregt, P., Contributors, S. 1. 0, 2019. SciPy 1.0–Fundamental Algorithms for Scientific Computing in Python. arXiv e-prints arXiv:1907.10121.

Virtanen, T., 2007. Monaural sound source separation by nonnegative matrix factorization with temporal continuity and sparseness criteria. IEEE Trans. Audio Speech Lang. Process. 15, 1066–1074. https://doi.org/10.1109/TASL.2006.885253.

Walt, S., van der, Colbert, S.C., Varoquaux, G., 2011. The NumPy array: a structure for efficient numerical computation. Comput. Sci. Eng. 13, 22–30. https://doi.org/10.1109/MCSE.2011.37.